\documentclass[prb,aps, preprint]{revtex4-2}

\usepackage[utf8]{inputenc}
\usepackage{amsmath}
\usepackage{graphicx}
\usepackage{xcolor}

\newcommand{\revmod}[1]{\textcolor{black}{#1}}
\newcommand{\revrevmod}[1]{\textcolor{black}{#1}}

\newcommand{\He}{{\hat H}_{\text{e}}}
\newcommand{\Hph}{{\hat H}_{\text{ph}}}
\newcommand{\Heph}{{\hat H}_{\text{e-ph}}}
\newcommand{\muHP}{\mu_{\textrm{H-P}}}
\newcommand{\muH}{\mu_{\textrm{H}}}
\newcommand{\muP}{\mu_{\textrm{P}}}
\newcommand{\muM}{\mu_{\textrm{M}}}

\newcommand{\CtHP}{C(t)_{\textrm{H-P}}}
\newcommand{\CtH}{C(t)_{\textrm{H}}}
\newcommand{\CtP}{C(t)_{\textrm{P}}}
\newcommand{\CtPA}{C(t)_{\textrm{PA}}}
\newcommand{\CtTL}{C(t)_{\textrm{TL}}}

\def\bra#1{\mathinner{\langle{#1}|}}
\def\ket#1{\mathinner{|{#1}\rangle}}
\def\braket#1{\mathinner{\langle{#1}\rangle}}

\begin{document}

\title{A General Charge Transport Picture for Organic Semiconductors with Nonlocal Electron-Phonon Couplings}

\author{Weitang Li}
 \affiliation{MOE Key Laboratory of Organic OptoElectronics and Molecular	
 Engineering, Department of Chemistry, Tsinghua University, Beijing 100084,	
 People's Republic of China }
\author{Jiajun Ren}
 \affiliation{MOE Key Laboratory of Organic OptoElectronics and Molecular	
 Engineering, Department of Chemistry, Tsinghua University, Beijing 100084,	
 People's Republic of China }
\author{Zhigang Shuai*}
%\email{zgshuai@mail.tsinghua.edu.cn}
 \affiliation{MOE Key Laboratory of Organic OptoElectronics and Molecular	
 Engineering, Department of Chemistry, Tsinghua University, Beijing 100084,	
 People's Republic of China }

\begin{abstract}
The nonlocal electron-phonon couplings in organic semiconductors 
responsible for the fluctuation of intermolecular transfer integrals has been the center of interest recently.
Several irreconcilable scenarios coexist for the description of the nonlocal electron-phonon coupling, 
such as phonon-assisted transport, transient localization, and band-like transport. 
Through a nearly exact numerical study for the carrier mobility of the Holstein-Peierls model using the matrix product states approach,
we locate the phonon-assisted transport, transient localization and band-like regimes as a function of the transfer integral ($V$) 
and the nonlocal electron-phonon couplings ($\Delta V$),
and their distinct transport behaviors are analyzed by carrier mobility, mean free path, optical conductivity and one-particle spectral function. We also identify an “intermediate regime” where none of the established pictures applies, and the generally perceived hopping regime is found to be at a very limited end in the proposed regime paradigm.
\end{abstract}

\maketitle

\section*{Introduction}

The last two decades have witnessed the rapid development of high-mobility
crystalline organic semiconductors~\cite{Podzorov04, Podzorov05, Takeya14}.
The first proposition of using Marcus’ semiclassical hopping model coupled with density functional theory to design high mobility molecules has been very successful and popular~\cite{Bredas02}.
It can be regarded as the strong local electron-phonon coupling (EPC) limit, which was later improved by considering quantum nuclear effect and delocalization effect~\cite{Nan09, Jiang16}, through which, isotope effect is found to be always negative and the dynamic disorder does not play appreciable role with some experimental supports~\cite{Jiang15, Wang10Multicale, Ren17, Ostroverkhova11}.
However, such local EPC picture is challenged by the recently established transient localization (TL) model which invokes nonlocal EPC~\cite{Fratini09, Fratini16, Fratini17, Sirringhaus19}.
In due course, the molecular design principles derived from TL such as suppressing intermolecular vibration,
which are quite different from the local picture, have proved successful in a number of experiments~\cite{Takeya16, Sirringhaus16}.
The applicability of the TL picture, nevertheless, is restricted to 
the regime of moderate transfer integral (electronic coupling) $V$ and strong nonlocal EPC.
Since EPC is a complicated many-body problem, it is highly desirable to present a general transport picture taking both local and nonlocal EPC into considerations in a rigorous way, instead of uncontrolled approximations.

Many recent efforts have been devoted
to developing approximate methods that are able to portray 
a broader parameter space, including the band-like (BL) conduction regime~\cite{Timothy20, Fratini2020PRR}.
Besides, unlike in the case of Holstein model in which it is beyond doubt
that local EPC represents an obstacle for carrier diffusion~\cite{Holstein1, Mishchenko15, Li20JPCL}, 
how nonlocal EPC affects mobility does not have a definitive answer and the interplay between the local and nonlocal EPC is unclear.
Early theoretical treatments for carrier mobility in crystalline organic semiconductors
such as the Munn-Silbey approach and the polaron transformation
often reach the conclusion that the nonlocal EPC
leads to phonon-assisted (PA) transport and enhances mobility~\cite{Silbey852, Hann041, Hann04Anisotropy},
in sheer contrast with the basic starting point of the TL scenario.
The findings of several numerical studies on the carrier mobility of organic semiconductors
also contradict with the TL theory~\cite{Wang10Multicale, Zhao12}. 
In principle, PA, TL and BL are all possible mechanisms
for charge transport with nonlocal EPC, valid at their respective parameter regimes,
yet a universal theoretical treatment for the role of nonlocal EPC is not available
due to the complex \revmod{many-body electron-phonon} interaction.

In this work, we present a nearly exact study of the charge transport mechanism in the Holstein-Peierls model
using the time-dependent finite temperature matrix product state (MPS) formalism~\cite{Schol11, Scholl19}.
By studying EPC effect on the carrier mobility, mean free path, optical conductivity and one-particle spectral function,
we have located the PA , TL and BL regimes simultaneously
on the transfer integral - nonlocal EPC strength plane.
We have also identified an intermediate regime where none of the existing pictures is truly applicable,
as a generalization of the hopping-band crossover in the Holstein model.

\section*{Results}
\subsection*{System Hamiltonian and Kubo Formula}

We take the following one-dimensional Holstien-Peierls model 
with nearest-neighbour interaction and periodic boundary condition:
\begin{equation}
\label{eq:H_hp}
\begin{aligned}
    \hat H & = \He + \Hph + \Heph \\
    \He & = -V \sum_n (c_{n+1}^\dagger c_n +  c_{n}^\dagger c_{n+1}) \\
    \Hph & = \sum_{n, m} \omega_m b^\dagger_{n, m} b_{n, m} + \sum_n \omega_\theta b^\dagger_{n, \theta} b_{n, \theta} \\
    \Heph & = \sum_{n, m} g_{m, \rm{I}} \omega_m (b^\dagger_{n, m} + b_{n, m}) c^\dagger_n c_n \\
    & +  \sum_{n} g_{\theta, \rm{II}} \omega_\theta (b^\dagger_{n, \theta} + b_{n, \theta}) 
    (c^\dagger_{n+1} c_{n} + c^\dagger_{n} c_{n+1})
\end{aligned}
\end{equation}
where $c^\dagger$ ($c$) and $b^\dagger$ ($b$) are the creation (annihilation) operator
for electron and phonon respectively, 
and $V$ is the intermolecular transfer integral.
\revmod{The electronic motion is limited to single-electron manifold.}
$\omega_m$ and $g_{m, \rm{I}}$ are the frequency and the dimensionless EPC constant of the $m$th intramolecular vibration mode.
$\omega_\theta$ and $g_{\theta, \rm{II}}$ are the intermolecular vibration counterparts.
$\hbar$ is set to 1.
The thermal  fluctuation $\Delta V$
is related to intermolecular coupling constant $g_{\theta, \rm{II}}$ by~\cite{Bredas09JPCC}:
% This should be a separate equation since \Delta V frequently appears
\begin{equation}
    \Delta V = g_{\theta, \rm{II}} \omega_\theta \sqrt{\coth{\frac{\omega_\theta}{2 k_B T} }}
\end{equation}
\revmod{
The one-dimensional model in Equation~(\ref{eq:H_hp}) is an approximation to realistic organic semiconductors,
which typically adopt two-dimensional transport network~\cite{Wang10Multicale, Fratini17, Gabriele19}.
In the Supplementary Fig.~4 we demonstrate that this approximation is valid for anisotropic materials
by comparing the simulated one-particle spectral function with experimental angle resolved ultraviolet photoemission spectra (ARUPS)~\cite{Bussolotti17},
and at the end of the section we go beyond the one-dimensional model to discuss the isotropy effect on the different transport regimes.
}

In order to elucidate how nonlocal EPC affects charge transport at different transport regimes
we focus on the role of transfer integral $V$ and nonlocal EPC constant $g_{\theta, \rm{II}}$ (or equivalently $\Delta V$ at a given $T$).
Other parameters are fixed throughout this paper unless otherwise specified
with values drawn from representative organic semiconductors.
In organic semiconductors, the most common values of $V$ and $\Delta V$
range from 10 meV to 150 meV and from 10 meV to 60 meV, respectively~\cite{Troisi18JPCC, Blumberger19, Fratini20}.
The intramolecular vibration frequency $\omega_m$ and local EPC constant $g_{m, \rm{I}}$ are taken from
our previous DFT calculations for rubrene
and the total $3N-6$ normal vibration modes are reduced to 4 effective modes~\cite{Jiang16, Li20JPCL},
namely $\omega_m=$ 26 meV, 124 meV, 167 meV and 198 meV,
with the corresponding dimensionless $g_{m, \rm{I}}$ 0.83, 0.26, 0.34 and 0.37, respectively.
The intermolecular vibration frequency $\omega_\theta$ is set to be 50 $\textrm{cm}^{-1}$ (6.2 meV)
as commonly adopted in literature~\cite{TROISI07, Fratini17, Timothy20}.
\revmod{
The system is translational-invariant and 
we do not consider the effect of both diagonal and off-diagonal static disorder here~\cite{BASSLER}.
}

The carrier mobility is obtained via the Kubo formula~\cite{Mahan00}:
\begin{equation}
\label{eq:kubo}
    \mu = \frac{1}{k_B T e_0} \int_{0}^{\infty} \braket{\hat j(t) \hat j(0)} dt
    = \frac{1}{k_B T e_0} \int_{0}^{\infty} C(t) dt
\end{equation}
where for the Holstein-Peierls Hamiltonian in Equation~(\ref{eq:H_hp}) 
the current operator $\hat j$ takes the form:
\begin{equation}
    \hat j = \frac{e_0 R}{i} \sum_n 
    \left [ -V + g_{\theta, \rm{II}} \omega_\theta (b^\dagger_{n, \theta} + b_{n, \theta}) \right ]
    (c_{n+1}^\dagger c_n  -  c_{n}^\dagger c_{n+1})
\end{equation}
Here $R$ is the intermolecular distance and is set to 7.2 Å as in the case of rubrene crystal.
\revmod{
Although we have treated the model as a closed system,
in our study the recurrence problem is not severe and $C(t)$ in general rapidly decays to nearly zero,
except when both $V$ and $\Delta V$ are small.
In such cases we resort to a more strict model with 10 modes in total
for the convergence of $C(t)$.
}
Lying at the heart of our calculation is the evaluation of the current-current correlation function $C(t)=\braket{\hat j(t) \hat j(0)}$,
which is achieved by the time-dependent MPS formalism~\cite{Schol11, Scholl19, Li20JPCL}.
In most of our simulations, 
the number of molecules in the periodic one-dimensional chain is 21 and the virtual bond dimension is 80.
More details on the model with 9 intramolecular modes
as well as numerical convergence check on system size and
MPS parameters are included in Supplementary Table~1, Supplementary Fig.~1 and Supplementary Fig.~2. 

\subsection*{Carrier Mobility}

\begin{figure}
\centering
  \includegraphics[width=.8\textwidth]{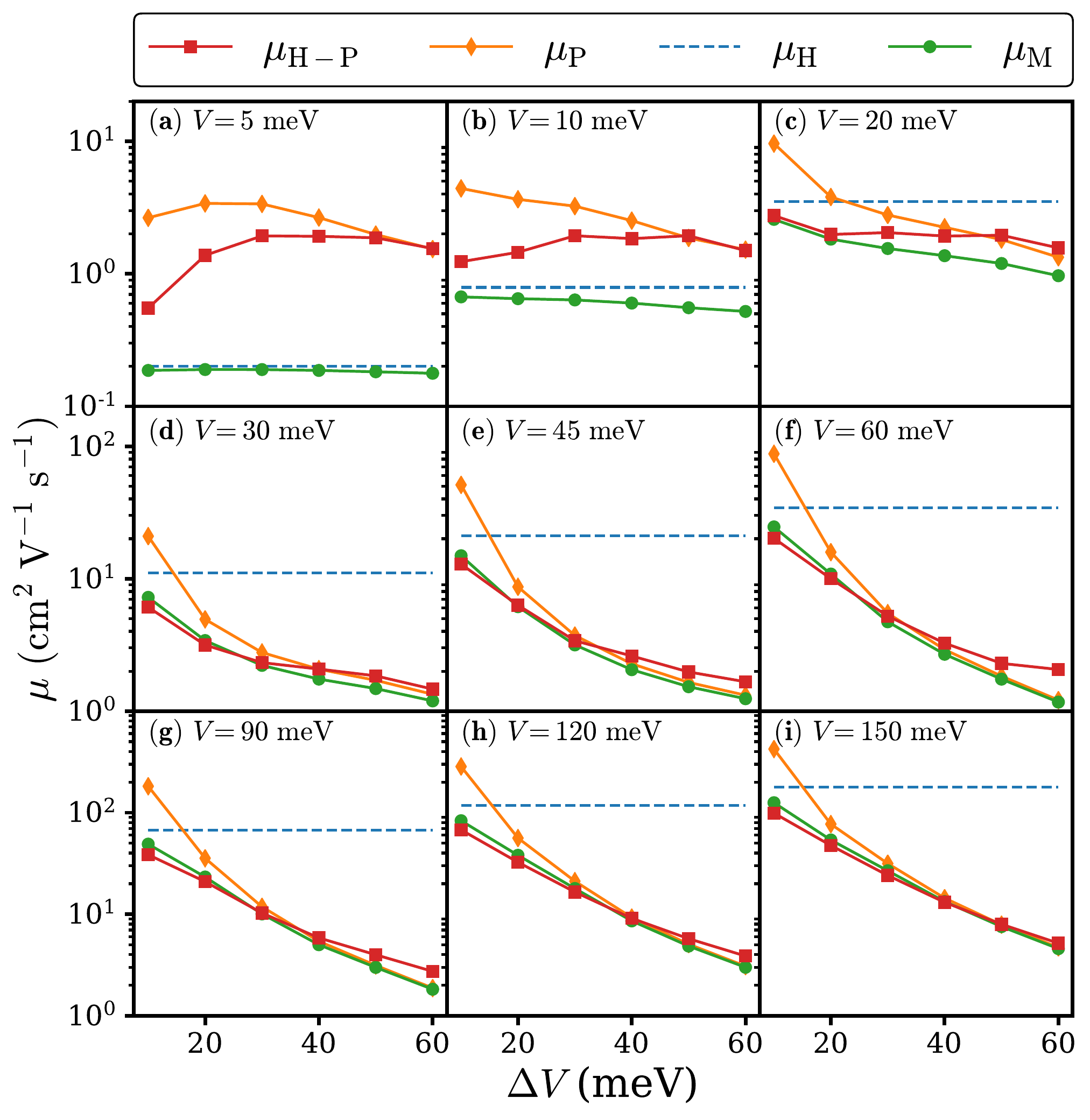}
  \caption{Carrier mobility at 300 K calculated based on the Holstein-Peierls model ($\muHP$), 
  Holstein model ($\muH$),
  Peierls model ($\mu_{\textrm{P}}$) and Matthiessen’s rule ($\muM$)
  with various transfer integral $V$ and transfer integral fluctuation $\Delta V$.
  From (a) to (i) the transfer integrals are 5, 10, 20, 30, 45, 60, 90, 120 and 150 meV respectively.
  Other parameters relevant to the carrier mobility such as local EPC constants are fixed with values taken from rubrene.
  The parameters for the Holstein (Peierls) model 
  is the same as that of Holstein-Peierls model except that the nonlocal (local) EPC is left out.}
  \label{fig:epc}
\end{figure}

Firstly, we analyze the role of local and nonlocal EPC on different parameter regimes
by comparing the mobility calculated based on the Holstein-Peierls model ($\muHP$) with 
the mobility calculated based on pure Holstein model ($\muH$) and pure Peierls model ($\muP$) at 
$300 \ \textrm{K}$.
%a bit nagging
%$\muH$ and $\muP$ are obtained by neglecting the nonlocal and local EPC of the Holstein-Peierls model respectively.
    We have also included the mobility derived from Matthiessen’s rule ($1/ \muM =1/\muH +1/\muP$),
presumably valid in the BL regime,
\revmod{
because in the BL regime both local and nonlocal EPC can be considered as independent scattering
sources and the total scattering rate of the wave-like electronic motion is the sum of both scattering rates.
}
The overall results are shown in Fig.~\ref{fig:epc}.
When $V = 5 \ \textrm{meV}$, $\muHP$ is generally higher than $\muH$, which implies that
nonlocal EPC is beneficial to charge transport. This is considered to be a signature of the
PA picture.
However, this behavior quickly vanishes as $V$ increases from 5 meV to 20 meV.
At the intermediate range of $V$, 
e.g. from 45 meV to 120 meV,
$\muHP$ could be higher than $\muP$ for larger $\Delta V$.
\revmod{
That local EPC could enhance instead of reduce mobility is quite counter-intuitive.
Such unusual behavior can be best understood by TL picture,
in which the quantum coherent interference responsible for Anderson localization
can be damaged or destroyed by local EPC as the dephasing noise.
The mechanism is studied in detail by means of open system dynamics 
for systems with static disorder~\cite{Cao13, Cao15}.
}
In addition, the ``band width narrowing" caused by local EPC could allow the carrier 
to thermally access much delocalized states~\cite{Timothy20}.
These lead to the local EPC enhanced mobility.
Upon further increasing $V$ to 150 meV, the TL scenario also becomes less significant.
Instead, it is found that $\muM$ coincides with $\muHP$ remarkably well,
serving as a piece of evidence for band-like transport.

\subsection*{Mean Free Path and Optical Conductivity}
\begin{figure}
\centering
  \includegraphics[width=.8\textwidth]{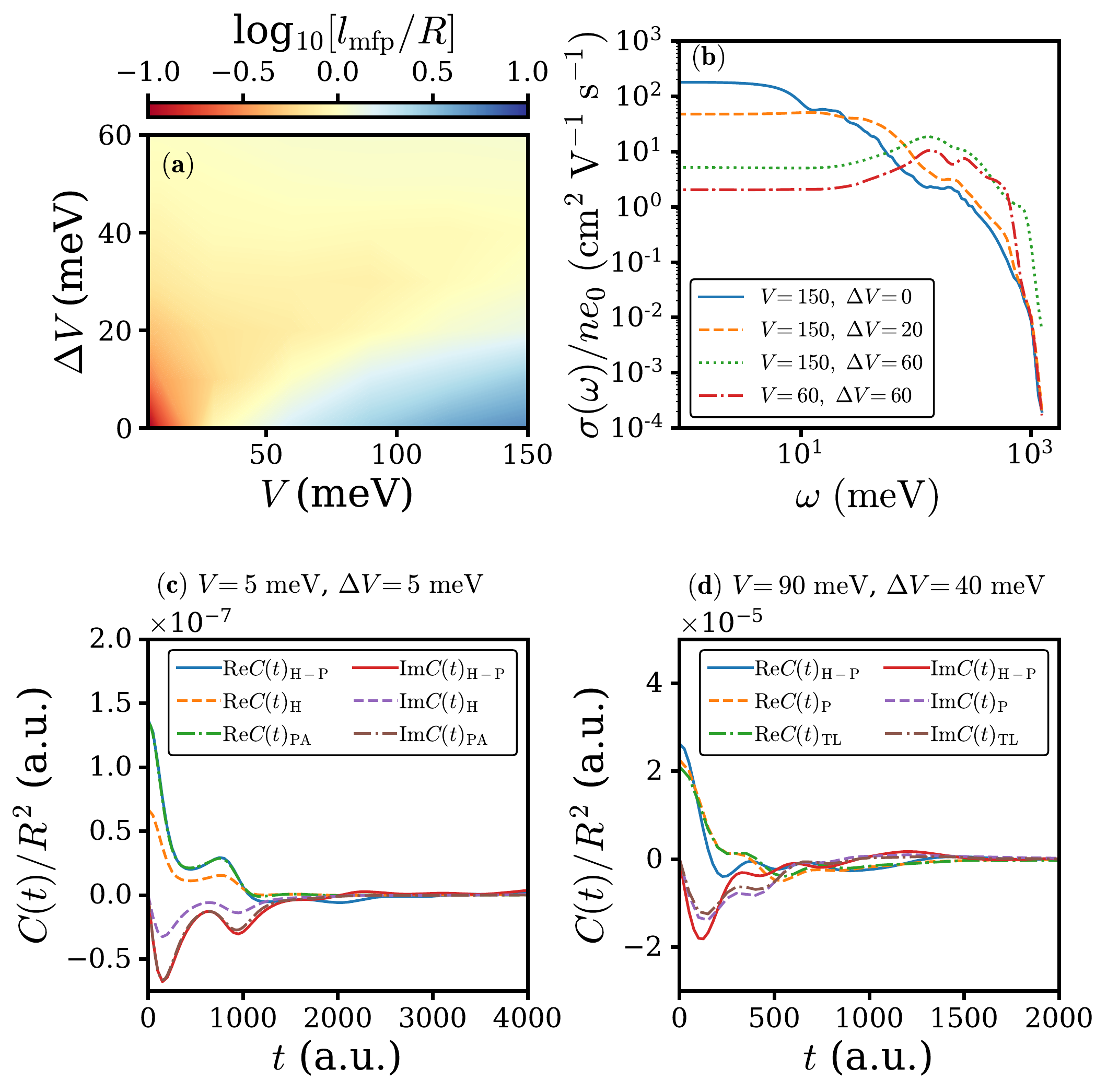}
  \caption{Further analysis of the transport regimes.
  (a) Carrier mean free path $l_{\textrm{mfp}}/R$ at 300 K for the Holstein-Peierls model at various transfer integral $V$ 
  and transfer integral fluctuation $\Delta V$. 
  In the blue region (bottom right) the carrier mean free path exceeds the lattice constant $R$.
  (b) Per carrier optical conductivity of the Holstein-Peierls model
  at various transfer integral $V$ 
  and transfer integral fluctuation $\Delta V$.
  \revmod{
  (c) and (d) Correlation functions obtained from our simulation with Holstein-Peierls model $\CtHP$, pure Holstein model $\CtH$ and pure Peiels model $\CtP$ as well as the correlation function obtained from phonon-assisted transport theory~\cite{Hann04Anisotropy} $\CtPA$ and transient localization theory~\cite{Fratini11} $C(t)_{\textrm{TL}}$ for (c) $V=5 \ \textrm{meV}$, $\Delta V = 5 \ \textrm{meV}$ and (d) $V=90 \ \textrm{meV}$, $\Delta V = 40 \ \textrm{meV}$.
  }
  }
  \label{fig:mfp}
\end{figure}

Although in the BL regime $\muM$ is expected to be a good approximation for $\muHP$,
$\muM \approx \muHP$ alone is not a sufficient condition for band-like transport,
and we additionally employ the Mott-Ioffe-Regel limit for the
determination of the BL regime.
The carrier mean free path $l_{\textrm{mfp}}$ is estimated as $l_{\textrm{mfp}}= v \tau$
with the group velocity $v$ and relaxation time $\tau$
evaluated by~\cite{Nenad19}:
\begin{equation}
\begin{aligned}
    v & =\sqrt{\langle {\hat j (0) \hat j(0)}} \rangle \Big / e_0 \\
    \tau& =\int_{0}^\infty \left | \frac{\rm{Re} C(t) } {\rm{Re} C(0) } \right | dt .
\end{aligned}
\end{equation}
And the calculated $l_{\textrm{mfp}}$ in the $(V, \Delta V)$ plane
at $300 \ \textrm{K}$ is shown in Fig.~\ref{fig:mfp}a.
The overall tendency of $l_{\textrm{mfp}}$
matches well with the carrier mobility of the Holstein-Peierls model $\muHP$ in Fig.~\ref{fig:epc}.
The region where $l_{\textrm{mfp}} > R$ is colored with blue in Fig.~\ref{fig:mfp}a
and it lies within the large $V$, small $\Delta V$ limit, in agreement with common perception.
Another possible criteria for BL conduction is the appearance of Drude-like peak in the 
per carrier optical conductivity:
\begin{equation}
\frac{\sigma(\omega)}{n e_0} = \frac{1-e^{-\omega/k_B T}}{\omega} \int_0^\infty C(t) e^{i \omega t} dt    
\end{equation}
which is illustrated in Fig.~\ref{fig:mfp}b.
In the case of $V = 150 \ \textrm{meV}$ without nonlocal EPC, a broad Drude-like peak appears near $\omega = 0$. 
Upon adding a small amount of nonlocal EPC, the Drude-like peak becomes invisible.
However, the optical conductivity is still significantly different from 
the $\Delta V = 60 \ \textrm{meV}$ cases,
in which a localization peak at $\omega  \approx 200 \ \textrm{meV} $ characteristic for the TL regime~\cite{Fratini2020PRR}
is present.

\revmod{
\subsection*{Connection with Semi-Analytical Theories}
The PA regime and the TL regime can be further confirmed by semi-analytical results.
In Fig.~\ref{fig:mfp}c we compare $\CtHP$ of our numerical simulation 
with $\CtPA$ of phonon-assisted charge transport theory~\cite{Hann04Anisotropy}:
\begin{equation}
\label{eq:pa}
\begin{aligned}
\mu & = \frac{e_0 R^2}{k_B T} \int_{-\infty} ^{\infty} [V^2 + (g_{\theta, \rm{II}} \omega_m)^2 \Phi_\theta(t)] e^{-\Gamma(t)} dt \\
\Gamma(t) & =2 \sum_m g_{m, \rm{I}}^2 [1 + 2N_m - \Phi_m(t)] + 4 g_{\theta, \rm{II}}^2 [1 + 2N_\theta - \Phi_m(t) ] \\
\Phi_m & = (1 + N_m) e^{-i\omega_m t} + N_m e^{i \omega_m t} \\
N_m & = \frac{1}{e^{\omega_m /k_BT}-1}
\end{aligned}
\end{equation}
The parameters are $V=5 \ \textrm{meV}$ and $\Delta V = 5 \ \textrm{meV}$.
For both real and imaginary part the two curves are in excellent agreement, 
and show significant increase with respect to the correlation function with only local EPC $\CtH$.
Thus we can confidently conclude that in this parameter regime 
the transport mechanism can be understood as phonon assisted transport. 
We note that the derivation of Equation~(\ref{eq:pa}) employs narrow-band approximation, 
which is valid in the hopping regime.
When $V=90 \ \textrm{meV}$ and $\Delta V = 40 \ \textrm{meV}$ shown in Fig.~\ref{fig:mfp}(d), 
the correlation function given by TL theory 
with relaxation time approximation~\cite{Fratini11} $\CtTL$ 
is in agreement with our calculation based on pure Peierls model $\CtP$. 
The observation implies that in this regime the TL theory can 
successfully account for the transport property of the pure Peierls model,
from which the TL theory is derived.
If the Holstein coupling is included, 
the correlation function $\CtHP$ exhibits significant difference 
from $\CtP$ and $\CtTL$, however, the integrated mobility turns out to be 
rather insensitive to Holstein coupling in this regime (Fig.~\ref{fig:epc}g). 
\revrevmod{We note that it is possible to integrate Holstein coupling in the transient localization
theory if the intramolecular vibration frequency is much smaller 
than the transfer integral~\cite{Fratini20, Troisi20AFM},
however such scheme is not employed in this work because in most cases $\omega_m$ is at the same order with $V$.}
We believe it is suitable to ascribe the $V=90 \ \textrm{meV}$ and $\Delta V = 40 \ \textrm{meV}$ case as 
TL, because although the TL theory with relaxation time approximation 
may not correctly produce the correlation function for realistic materials with Holstein coupling, 
the picture provided by the theory serves as a nice starting point for further analysis.
}

\begin{figure}[h]
  \includegraphics[width=\textwidth]{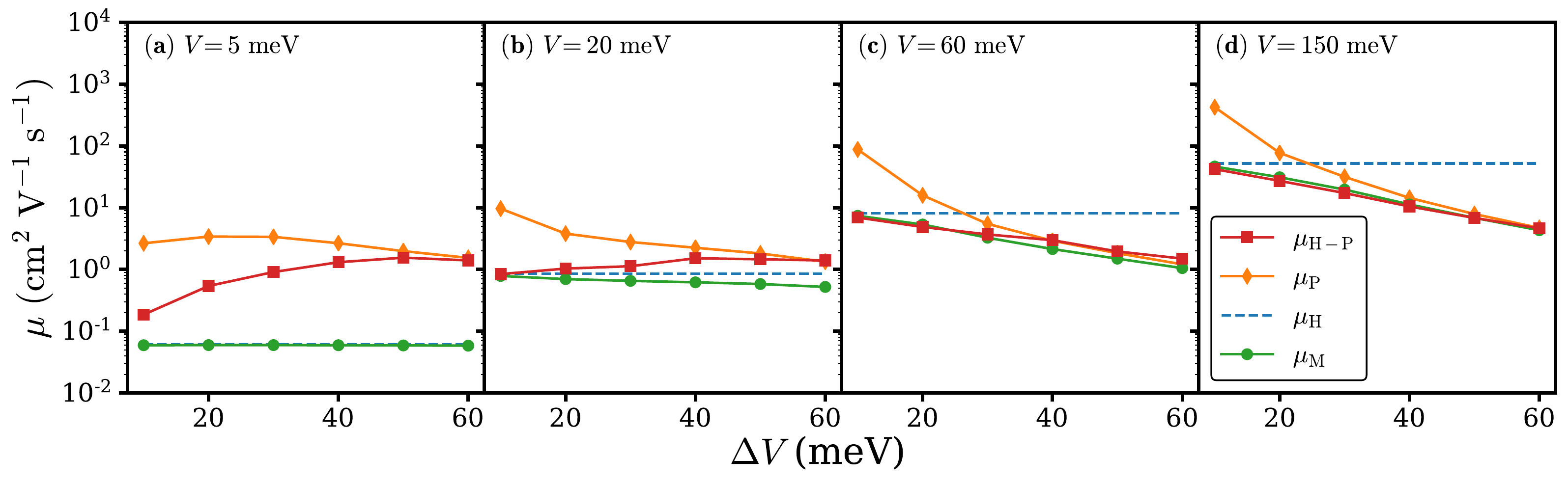}
  \caption{\revmod{
  Carrier mobility at 300 K calculated based on the Holstein-Peierls model ($\muHP$), 
  Holstein model ($\muH$),
  Peierls model ($\mu_{\textrm{P}}$) and Matthiessen’s rule ($\muM$)
  with enlarged local EPC at various transfer integral $V$ 
  and transfer integral fluctuation $\Delta V$.
    From (a) to (d) the transfer integrals are 5, 20, 60, and 150 meV respectively.
  }
  }
  \label{fig:si-epc}
\end{figure}

\begin{figure}[h]
\centering
  \includegraphics[width=.4\textwidth]{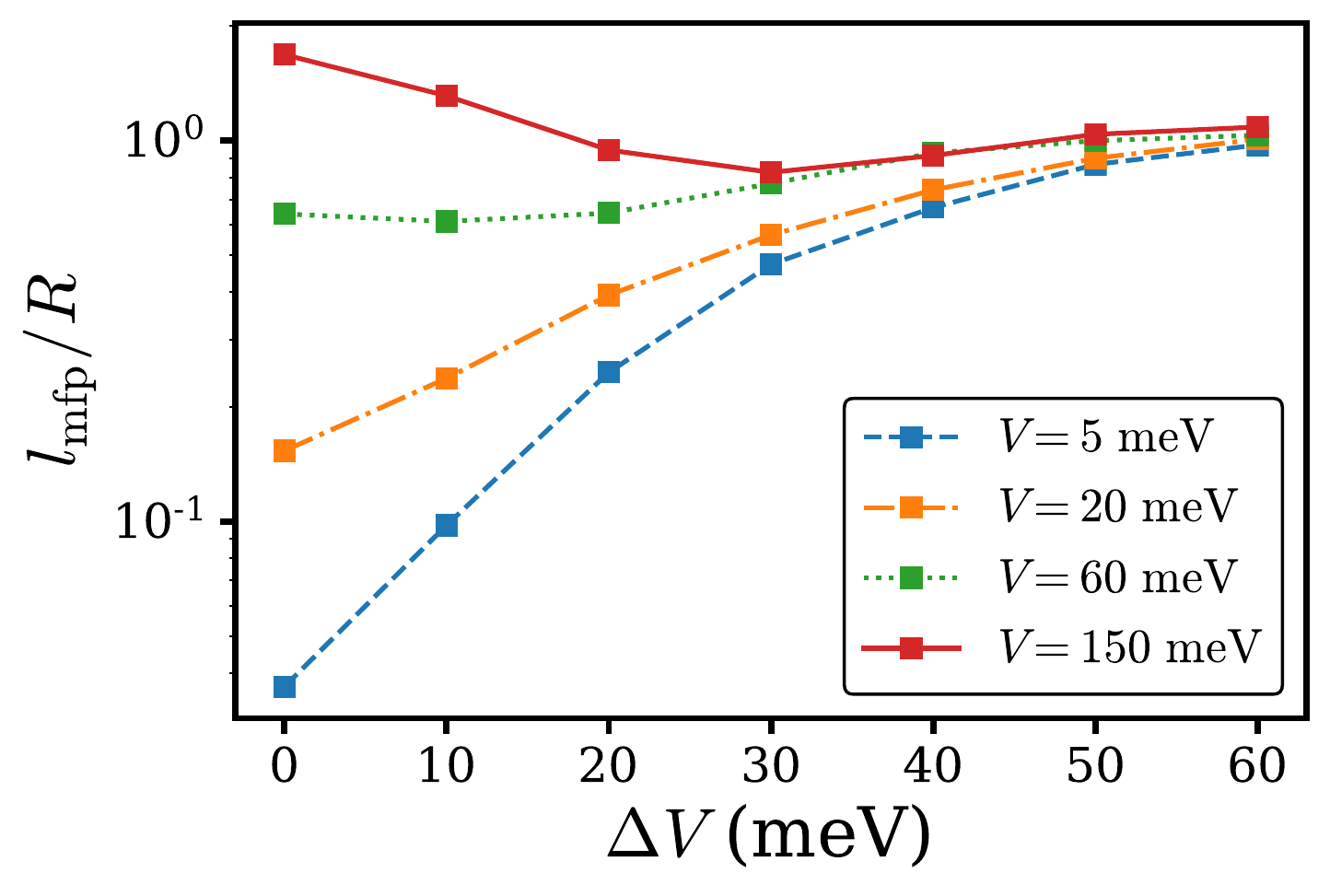}
  \caption{ 
  \revmod{
  Carrier mean free path $l_{\textrm{mfp}}/R$ at 300 K for the Holstein-Peierls model with enlarged local EPC
  at various transfer integral $V$ 
  and transfer integral fluctuation $\Delta V$.
  }
  }
  \label{fig:si-mfp}
\end{figure}

\revmod{
\subsection*{Effect of Local EPC Strength}
In order to investigate how local EPC strength will affect the results in Fig.~\ref{fig:epc},
we have further calculated the carrier mobility of the Holstein-Peierls model with stronger local EPC.
More specifically, the values of $g_{m, \rm{I}}$ are multiplied by $\sqrt{2}$ 
so that the respective reorganization energies $g_{m, \rm{I}}^2 \omega_m$ are doubled.
The results are illustrated in Fig.~\ref{fig:si-epc}.
In the small $V$ limit shown in Fig.~\ref{fig:si-epc}(a), the PA mechanism prevails,
in agreement with the results in Fig.~\ref{fig:epc}.
However, from Fig.~\ref{fig:si-epc}(b) it can be seen that with enlarged local EPC strength 
the PA picture remains valid even if $V$ becomes as large as $20 \ \textrm{meV}$,
in contrast to the $V = 20 \ \textrm{meV}$ results presented in Fig.~\ref{fig:epc}, 
indicating the expansion of the PA region.
Accordingly, the TL region is diminished,  as can be inferred from Fig.~\ref{fig:si-epc}(c) and (d) 
by noting that the parameter space in which $\muHP \ge \muP$ is satisfied is smaller than that of Fig.~\ref{fig:epc}.
In Fig.~\ref{fig:si-mfp} we show the carrier mean free path $l_{\textrm{mfp}}/R$ with increased local EPC strength.
When $V$ is relatively large and $\Delta V$ is relatively small, namely in the BL regime,
$l_{\textrm{mfp}}/R$ with increased local EPC strength is generally smaller than that of the original local EPC strength.
This observation implies that the BL region in the $(V, \Delta V)$ plane 
moves toward the even larger $V$ area ($V > 150 \ \textrm{meV}$).
Our findings are in agreement with physical instinct because in the large local EPC limit
the hopping mechanism dominates and Equation~(\ref{eq:pa}) is a good approximation for mobility.
\revrevmod{
Another factor that might affect charge transport is the distribution of intramolecular vibration frequencies 
with fixed total reorganization energy.  
The problem is equivalent to the isotope effect problem and recent studies on the rubrene molecule concluded
negative isotope effect~\cite{Jiang16,Ren17,Li20JPCL}.
}
}

\subsection*{One-Particle Spectral Function}

\begin{figure}
\centering
  \includegraphics[width=.8\textwidth]{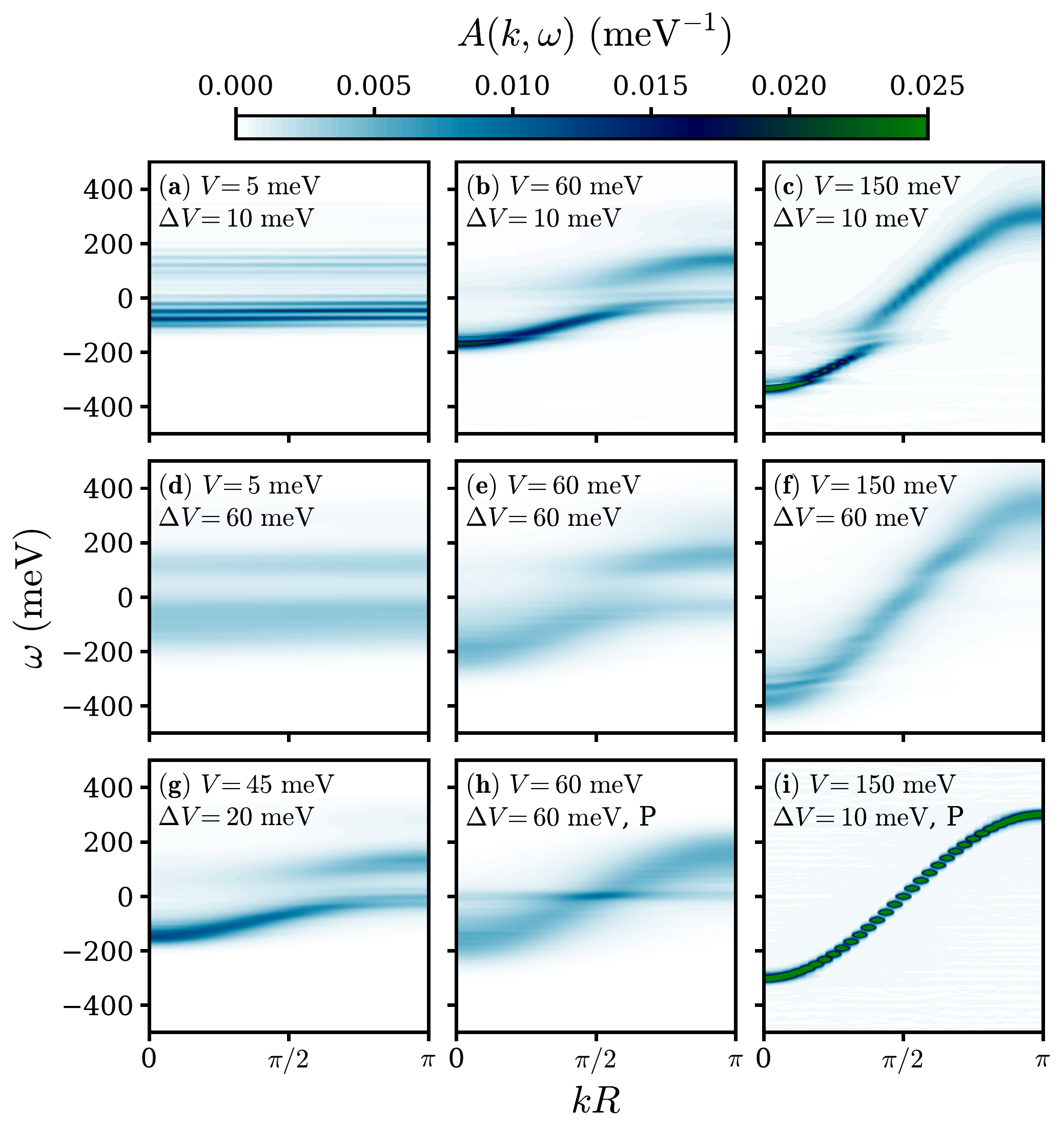}
  \caption{Spectral function at 300 K for the Holstein-Peierls model [(a) to (g)] and the pure Peierls model [(h) and (i)].
  Panel (h) and (i) has the same parameter as panel (e) and (c) respectively except that 
  in Panel (h) and (i) the local EPC is left out.
  }
  \label{fig:akw}
\end{figure}

To further analyze the charge transport properties in the regimes implied by Fig.~\ref{fig:epc} and Fig.~\ref{fig:mfp},
we calculated the momentum resolved one-particle spectral function: 
\begin{equation}
    A(k, \omega) = \frac{1}{N\pi} \sum^N_{mn} e^{ikR(m-n)} \int_0^\infty \braket{c_m(t) c_n^\dagger(0)} e^{i \omega t} dt 
\end{equation}
for nine sets of representative parameters at 300 K 
in Fig.~\ref{fig:akw}.
A Lorentzian broadening with $\eta=5 \ \textrm{meV}$ is applied for a smooth spectra.
When $V = 5 \ \textrm{meV}$ and $\Delta V = 10 \ \textrm{meV}$ (Fig.~\ref{fig:akw}a),
the spectral function exhibits dispersionless bound states
separated by the intramolecular vibration frequencies $\omega_m$,
which marks the formation of small polaron.
On the contrary, when $V = 150 \ \textrm{meV}$ and $\Delta V = 10 \ \textrm{meV}$ (Fig.~\ref{fig:akw}c)
an intense quasiparticle peak is observed near $k=0$ and the overall shape of the spectra resembles
the dispersion of free electron $E(k) = -2V\cos{kR}$.
In the TL regime represented by $V = 60 \ \textrm{meV}$ and $\Delta V = 60 \ \textrm{meV}$ (Fig.~\ref{fig:akw}e)
the signature of either small polaron or delocalized state is almost completely smeared out.
In combination with the limited carrier mean free path in this regime, 
it can be deduced that the charge carrier is localized by nonlocal EPC instead of local EPC.
The same ``blurred'' spectral function is observed for other sets of typical parameters in the TL regime (Fig.~\ref{fig:akw}d, f).
With moderate $V$ and $\Delta V$ shown in Fig.~\ref{fig:akw}g,
the spectral function exhibits none of the typical features described above.
Namely, although the spectral function does not manifest the formation of small polaron or delocalized states,
the peak intensity is still strong enough to be discernible from the TL regimes cases (Fig.~\ref{fig:akw}d, e, f).
In the absence of the local EPC (Fig.~\ref{fig:akw}h, i), the quasiparticle peak has more intensity,
implying the disruption of quantum coherent with the addition of local EPC.
Combined with the mobility data shown in Fig.~\ref{fig:epc},
it can be inferred that
in the TL regime, the effect of the disruption is to
alleviate the localization caused by nonlocal EPC, leading to increased mobility (Fig.~\ref{fig:akw}e, h),
while in the BL regime, on the contrary, the disruption scatters charge carrier and reduces mobility (Fig.~\ref{fig:akw}c, i).
The horizontal peak at the center of the band in Fig.~\ref{fig:akw}h
is a result of the pure nonlocal EPC in the Peierls model
and is expected to vanish upon the inclusion of infinitesimal local EPC~\cite{Cohen76}.

\subsection*{The Isotropy Effect}
\label{sec:isotropy}
\begin{figure}[h]
\centering
  \includegraphics[width=\textwidth]{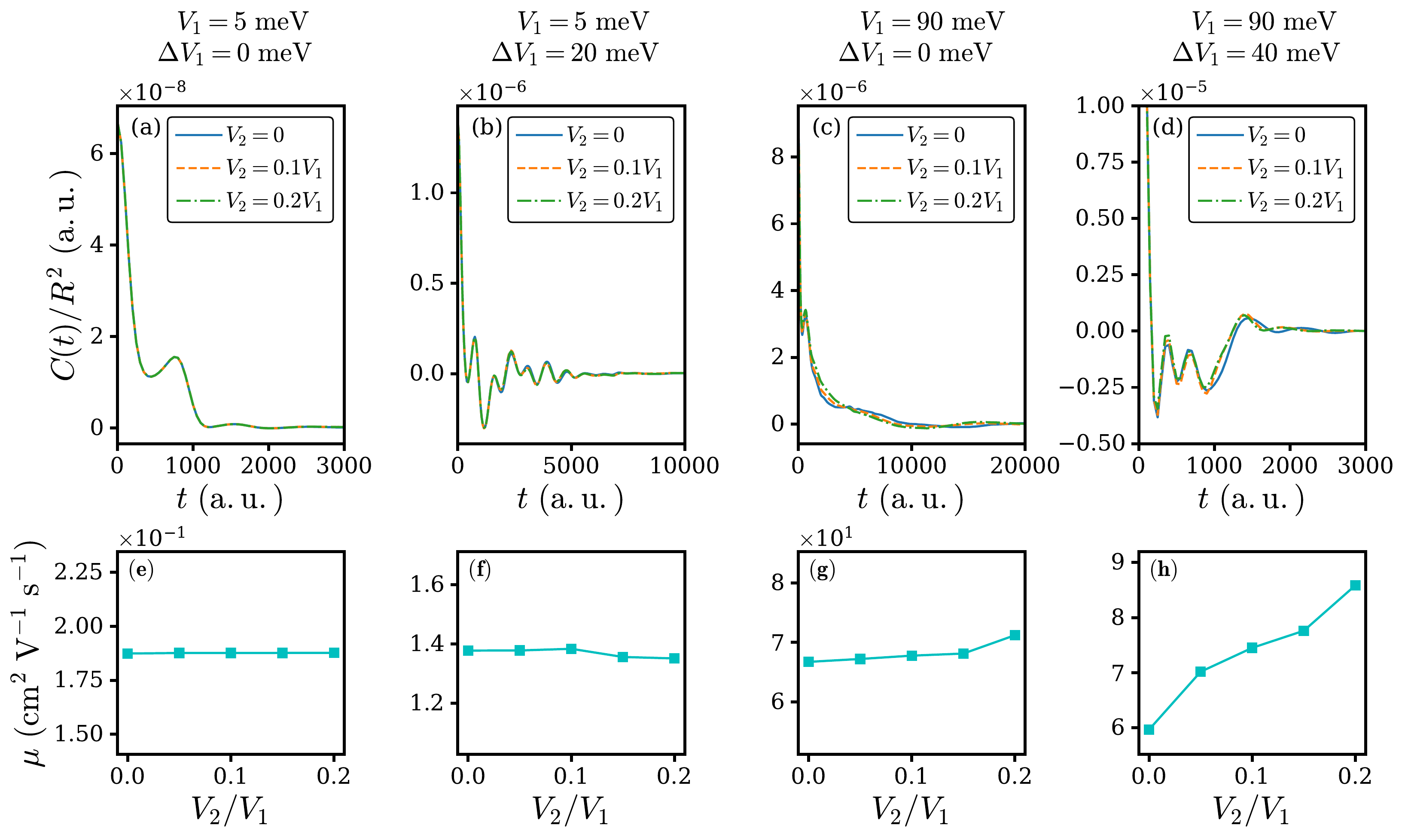}
  \caption{
  \revmod{The isotropy effect for several typical values of $V_1$ and $\Delta V_1$. 
  (a) to (d) are the correlation functions $C(t)$ and (e) to (h) are the corresponding mobilities $\mu$. 
  For (a) and (e), $V_1=5 \ \textrm{meV}$ and $\Delta V_1=0 \ \textrm{meV}$; 
  For (b) and (f), $V_1=5 \ \textrm{meV}$ and $\Delta V_1=20 \ \textrm{meV}$; 
  For (c) and (g), $V_1=90 \ \textrm{meV}$ and $\Delta V_1=0 \ \textrm{meV}$; 
  For (d) and (h), $V_1=90 \ \textrm{meV}$ and $\Delta V_1=40 \ \textrm{meV}$.
  }
  }
  \label{fig:isotropy}
\end{figure}

\revmod{
In a number of recent works it is established that dimensionality plays an indispensable role in the charge transport process, especially when dynamic disorder 
is taken into consideration~\cite{Wang10Multicale, Bobbert12, Fratini17}.
To study the isotropy effect beyond the one-dimensional model,
we employ a quasi-two-dimensional ladder Holstein-Peierls Hamiltonian:
\begin{equation}
\label{eq:isotropy}
\begin{aligned}
    \hat H & = \He + \Hph + \Heph \\
    \He & = -V_1 \sum_{l=1,2}\sum_n (c_{l, n+1}^\dagger c_{l, n} +  c_{l, n}^\dagger c_{l, n+1})
            -V_2 \sum_n (c_{0, n}^\dagger c_{1, n} +  c_{1, n}^\dagger c_{0, n}) \\
    \Hph & = \sum_{l, n, m} \omega_m b^\dagger_{l, n, m} b_{l, n, m} + \sum_{l, n} \omega_\theta b^\dagger_{l, n, \theta} b_{l, n, \theta} \\
    \Heph & = \sum_{l, n, m} g_{m, \rm{I}} \omega_m (b^\dagger_{l, n, m} + b_{l, n, m}) c^\dagger_{l, n} c_{l, n} \\
    & +  \sum_{l, n} g_{\theta, \rm{II}} \omega_\theta (b^\dagger_{l, n, \theta} + b_{l, n, \theta}) 
    (c^\dagger_{l, n+1} c_{l, n} + c^\dagger_{l, n} c_{l, n+1})
\end{aligned}
\end{equation}
here $V_1$ and $V_2$ represent the electronic coupling 
at the high-mobility direction and the low-mobility direction respectively. 
The intermolecular vibration at the $V_2$ direction is neglected for simplicity. 
The setup, while still approximate compared to a full-fledged two-dimensional model, 
is believed to be reasonable for anisotropic materials ($V_2 \ll V_1$) 
and can at least partially capture the dimensionality effect.
In Fig.~\ref{fig:isotropy} we present the computed correlation function $C(t)$ and mobility $\mu$ 
based on the model for several typical values of $V_1$ and $\Delta V_1$. 
In the hopping limit shown in Fig~\ref{fig:isotropy}a and Fig~\ref{fig:isotropy}e, 
it is found that carrier mobility is irrelevant to the isotropy effect, 
because in this limit $V_2$ does not affect the hopping process at $V_1$ direction. 
In the phonon-assisted transport regime shown in Fig.~\ref{fig:isotropy}b and Fig.~\ref{fig:isotropy}f, 
$\mu$ is rather insensitive to isotropy effect. 
In the band-like regime shown in Fig.~\ref{fig:isotropy}c and Fig.~\ref{fig:isotropy}g, 
we find that isotropy effect tends to slightly increase mobility. 
Lastly, in the transient localization regime shown in Fig.~\ref{fig:isotropy}d and Fig.~\ref{fig:isotropy}h, 
it is observed that carrier mobility is susceptible to the isotropy effect.
By increasing $V_2$ from 0 to $0.2 V_1$, the mobility increases by about 40\%. 
Such increase may appear difficult to understand as $C(t)$ in Fig.~\ref{fig:isotropy}d does not seem to vary much.
This is because transient localization implies that
during the integration of $C(t)$ the positive and the negative part of $C(t)$ are cancelled out, 
and thus minor changes in $C(t)$ will result in a big difference in mobility.
Our result is generally in agreement with previous literatures~\cite{Wang10Multicale, Fratini17}.
Based on these findings we can conclude that when the isotropy of the system is increased, 
the band region tend to expand while the transient localization regime would tend to diminish~\cite{Fratini2020PRR}.
Physically, the first conclusion can be understood by enlarged bandwidth in two dimension and the second conclusion can be understood by considering that Anderson localization length for two dimension is larger than that in one dimension~\cite{Patrick85}.
}

\subsection*{General Charge Transport Regime Diagram}

\begin{figure}
\centering
  \includegraphics[width=.7\textwidth]{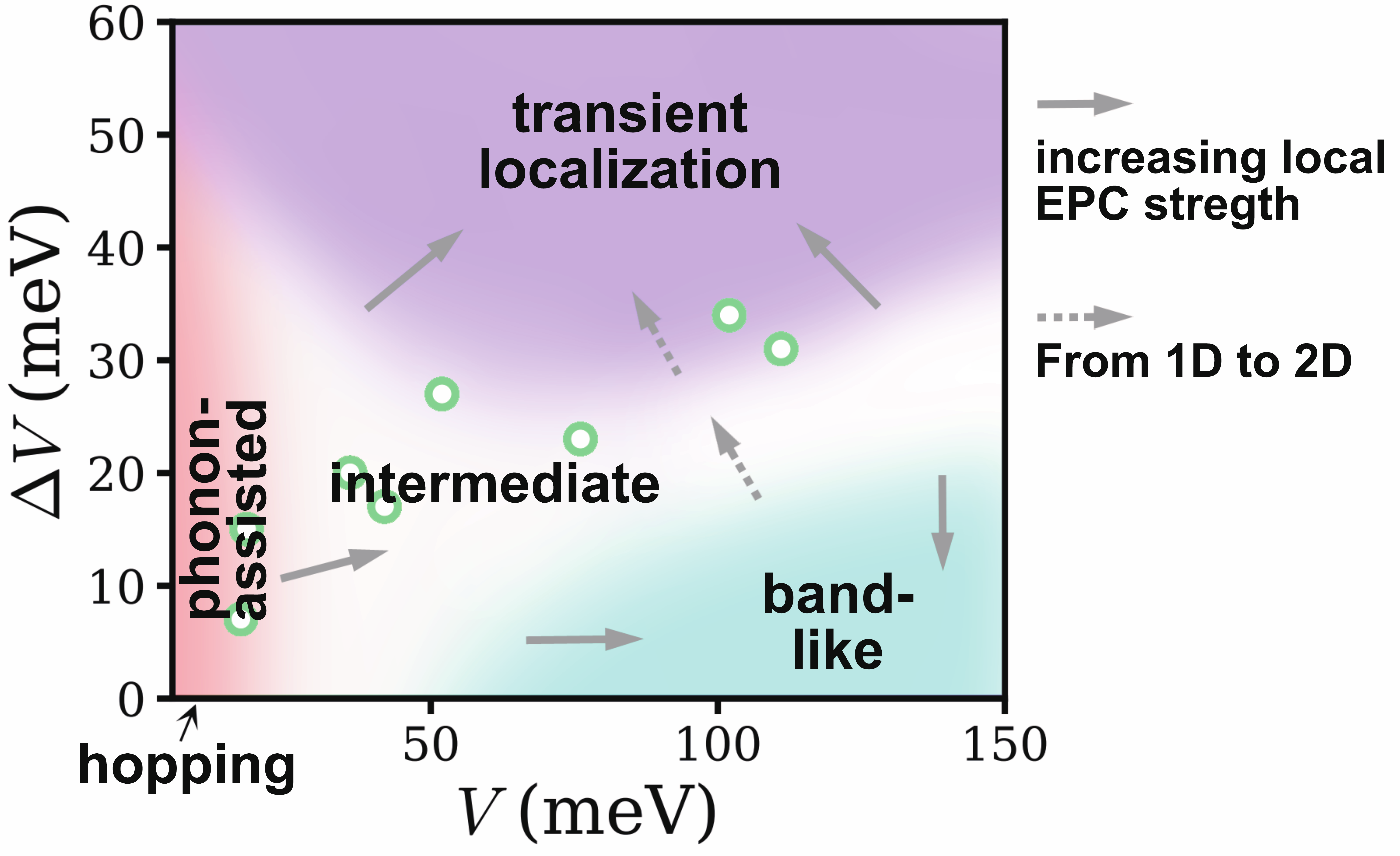}
  \caption{A schematic ``regime diagram" showing the
   phonon-assisted transport (PA) regime, transient localization (TL) regime, band-like (BL) regime
   and intermediate regime on the $(V, \Delta V)$ plane
  for the carrier mobility of the Holstein-Peierls model.
  The hopping regime is achieved in the $\Delta V = 0$ limit of the PA regime.
  Grey solid arrows show qualitatively the shift of the boundaries when local EPC increases.
  \revmod{
  Grey dashed arrows show qualitatively the shift of the boundaries when transport network
  changes from one-dimension to quasi-two-dimension or equivalently when electronic coupling isotropy increases.
  }
  The green dots represent the $V$ and $\Delta V$ of several common organic semiconductors.}
  \label{fig:phase_diagram}
\end{figure}

Based on the EPC effect on the carrier mobility, the mean free path, the optical conductivity and the one-particle spectral function,
we are able to sketch a schematic ``regime diagram" for the charge transport mechanisms
as shown in Fig.~\ref{fig:phase_diagram}.
\revmod{
The PA regime is determined by $\muHP > \muH$, short  $l_{\textrm{mfp}}$, 
$C(t)_{\textrm{H-P}} \approx C(t)_{\textrm{PA}}$ and narrow bound state states in the spectral function.
The TL regime is determined by $\muHP \ge \muP$, intermediate $l_{\textrm{mfp}}$,
localization peak in optical conductivity,
$C(t)_{\textrm{P}} \approx C(t)_{\textrm{TL}}$
and a ``smeared out'' spectral function.
The BL regime is determined by  $\muHP \approx \muM$, 
large $l_{\textrm{mfp}}$, Drude-like peak in optical conductivity
and sharp quasiparticle peak in the spectral function.
In Fig.~\ref{fig:phase_diagram} we use $\muHP > \muH$, $\muHP \ge \muP$ and
$l_{\textrm{mfp}} > R$ for the boundaries of the PA regime, TL regime and BL regime respectively,
and using other indicators such as the appearance of Drude-like peak for the BL regime
may shift the boundaries to some extent but the general picture remains intact.
}
On this $(V, \Delta V)$ plane we are also able to identify an ``intermediate" regime that lies among
the PA regime, TL regime and BL regime. 
In this regime, $\muHP$ is significantly lower than both $\muH$ and $\muP$,
and the carrier mean free path  is still less than the lattice constant, forbidding the band description.
In fact, for the pure Holstein model case ($\Delta V = 0$), 
the intermediate regime simply degenerates into the canonical hopping-band crossover.
\revmod{
The crossover from the BL regime to the TL regime has also been reported
by introducing transient localization correction to band transport~\cite{Fratini2020PRR}.
The grey solid arrows and grey dashed arrows in Fig.~\ref{fig:phase_diagram} indicate the shift of the boundaries 
upon increasing local EPC and increasing electronic coupling isotropy respectively,
based on Fig.~\ref{fig:si-epc}, Fig.~\ref{fig:si-mfp} and Fig.~\ref{fig:isotropy}.
}
To provide a rough intuition of the distribution of
parameters for realistic organic semiconductors
on this $(V, \Delta V)$ plane, in Fig.~\ref{fig:phase_diagram}
we have also marked the value of $V$ and $\Delta V$ for several
common organic semiconductors
using reported values from recent literature~\cite{Blumberger19}.
These materials are pMSB, pyrene, naphthalene, perylene,
anthracene, DATT, rubrene and pentacene from left to right.
We note that the colors at the location of the markers do not 
represent a prediction of the charge transport mechanism for the corresponding materials
because the materials do not necessarily share the same local EPC coupling strength,
transport network and intermolecular vibration frequency with the parameters used in this work.

\section*{Discussion}
In this work, we present a nearly exact theoretical study of the carrier mobility in Holstein-Peierls model
with parameters relevant to organic semiconductors.
By carefully investigating the effect of both local and nonlocal EPCs on the carrier mobility $\mu$,
mean free path $l_{\textrm{mfp}}$, 
per carrier optical conductivity $\frac{\sigma(\omega)}{n e_0}$
and one-particle spectral function $A(k, \omega)$,
we are able to identify the PA regime, TL regime and BL regime on the $(V, \Delta V)$ plane.
The PA regime features $\muHP > \muH$, short $l_{\textrm{mfp}}$,
and narrow bound states in the spectral function.
The TL regime features $\muHP \ge \muP$, intermediate $l_{\textrm{mfp}}$,
localization peak in optical conductivity
and a ``smeared out'' spectral function.
And the BL regime features $\muHP \approx \muM$, large $l_{\textrm{mfp}}$,
Drude-like peak in optical conductivity
and sharp quasiparticle peak in the spectral function.
The semiclassical Marcus hopping regime is found to be around the corner of small $V$ and $\Delta V$.
Furthermore, some of the parameters in the $(V, \Delta V)$ plane are recognized to lie
in an intermediate regime that does not exhibit the typical features described above,
and this regime can be considered as a generalization of the hopping-band crossover regime
in the Holstein model.
\revmod{
We find that when increasing local EPC strength, the PA regime will expand 
while the TL and BL regime will diminish.
When going from one-dimension to quasi-two-dimension, the TL regime will diminish and the BL regime will expand.
}
It should be noted that the localization effect due to static disorder is not considered here, 
which deserves further investigation.

\section*{Methods}

\subsection*{Matrix Product States}
The evaluation of the current-current correlation function $C(t)=\braket{\hat j(t) \hat j(0)}$
is performed by time dependent matrix product states through imaginary and real time propagation.
\revmod{
The matrix product states method represent the wavefunction of many-body system as the product of
a series of matrices~\cite{Schol11}:
\begin{equation}
\label{eq:mps}
    \ket{\Psi}  = \sum_{\{a\},\{\sigma\}}
     A^{\sigma_1}_{a_1} A^{\sigma_2}_{a_1a_2} \cdots
           A^{\sigma_N}_{a_{N-1}}  \ket{ \sigma_1\sigma_2\cdots\sigma_N }
\end{equation}
$\ket{\sigma_i}$ is the basis for each degree of freedom.
$A^{\sigma_i}_{a_{i-1}a_{i}}$ are matrices in the chain connected by indices $a_i$. $\{ \cdot \}$ in the summation represents the contraction of the respective connected indices, 
and $N$ is the total number of degrees of freedom (DOFs) in the system.
The dimension of $a_i$ is called (virtual) bond dimension,
while the dimension of $\sigma_i$ is called physical bond dimension.
In principle, the time dependent algorithms for MPS~\cite{Scholl19} is able to solve the time dependent Schr\"odinger equation in an exact manner if the bond dimension is infinite. In practice, the accuracy of the method can be systematically improved by using a larger bond dimension, until convergence of interested physical observables within arbitrary convergence criteria.
}

\subsection*{Finite Temperature Algorithm}
The finite temperature effect is taken into account through thermal field dynamics, 
also known as the purification method~\cite{White05ft,Schol11}.
The thermal equilibrium density matrix of any mixed state in physical space $P$ can be expressed as a partial trace over an enlarged Hilbert space $P\otimes Q$, where $Q$ is an auxiliary space chosen to be a copy of $P$.
The thermal equilibrium density operator can then expressed
as a partial trace of the pure state $\Psi_\beta$ in the enlarged Hilbert space over the $Q$ space:
\begin{equation}
    \hat \rho_\beta = \frac{e^{-\beta \hat{H}}}{Z} = \frac{\rm{Tr}_Q\ket{\Psi_{\beta}}\bra{ \Psi_{\beta}}}
    {\rm{Tr}_{PQ}\ket{\Psi_{\beta}}\bra{ \Psi_{\beta}}}
\end{equation}
and the pure state $\ket{ \Psi_\beta }$
represented as an MPS is obtained by the imaginary time propagation
from the locally maximally entangled state $\ket{I}=\sum_i \ket{i}_P \ket{i}_Q$ to $\beta/2$ 
\revmod{in the one electron manifold}:
\begin{equation}
    \ket{\Psi_\beta} = e^{-\beta \hat H / 2} \ket{I} .
\end{equation}

To calculate $C(t)$, 
$\ket{\Psi_\beta}$ and $\hat j (0) \ket{\Psi_\beta}$ 
are propagated in real time to obtain
$e^{-i \hat H t} \ket{\Psi_\beta}$ and $e^{-i \hat H t} \hat j (0) \ket{\Psi_\beta}$ and then $C(t)$ is calculated by:
\begin{equation}
    C(t) = \bra{\Psi_\beta} e^{i \hat H t} \hat j (0) e^{-i \hat H t} \hat j (0) \ket{\Psi_\beta} / Z .
\end{equation}
Here the current operator $\hat j (0)$ is 
represented as an MPO and inner-product for $\ket{\Psi_\beta}$ includes tracing over 
both $P$ space and $Q$ space.
The construction of the MPOs is performed in an automatic and optimal fashion 
through our recently proposed algorithm~\cite{Ren20}.
\revmod{
Note that different from the simulation of diffusion dynamics, the initial state of the formulation does not require electronic excitation from the zero electron manifold.
}
In principle, both imaginary and real time propagation can be carried
out by any time evolution methods available to matrix product states~\cite{Scholl19}.
In this work, we use the time-dependent variational principle based projector splitting time evolution scheme~\cite{Haeg11, Haeg16}, 
which is found to be relatively efficient and accurate combined with graphic processing unit (GPU) in our recent work~\cite{li2020numerical}.

\section*{Acknowledgements}
This work is supported by the National Natural Science Foundation of China (NSFC) through the project ``Science CEnter for Luminescence from Molecular Aggregates (SCELMA)'' Grant Number 21788102, as well as by the Ministry of Science and Technology of China through the National Key R\&D Plan Grant Number 2017YFA0204501. J.R. is also supported by the NSFC via Grant Number 22003029.

\section*{Author contributions}
Z. S. conceived and supervised the study. 
W. L. performed the numerical calculations, analyzed the data and wrote the manuscript
with the help of J. R.

\section*{Data Availability}
The data generated in this study has been deposited in Zenodo with DOI 10.5281/zenodo.5009584.

\section*{Code Availability}
The computer code for the MPS algorithms used in this work is available publicly
via https://github.com/shuaigroup/Renormalizer~\cite{Reno21}.

\section*{Competing Interests}
The authors declare no competing interests.

\end{document}